\documentstyle[prd,aps]{revtex}

\def\laq{\ \raise 0.4ex\hbox{$<$}\kern -0.8em\lower 0.62
ex\hbox{$\sim$}\ }
\def\gaq{\ \raise 0.4ex\hbox{$>$}\kern -0.7em\lower 0.62
ex\hbox{$\sim$}\ }
\def\half{\hbox{\magstep{-1}$\frac{1}{2}$}}

\renewcommand\({\left(}
\renewcommand\){\right)}
\renewcommand\[{\left[}
\renewcommand\]{\right]}

\def\lsim{\raise 0.4ex\hbox{$<$}\kern -0.8em\lower 0.62
ex\hbox{$\sim$}}
\def\gsim{\raise 0.4ex\hbox{$>$}\kern -0.7em\lower 0.62
ex\hbox{$\sim$}}
\newcommand{\mpl}{M_{\rm Pl}}

\newcommand{\ogw}{\Omega_{\rm gw}}
\newcommand{\hogw}{h_0^2\Omega_{\rm gw}}

\newcommand{\fp}{f_{\rm peak}}

\newcommand\ee{\end{equation}}
\newcommand\be{\begin{equation}}
\newcommand\eea{\end{eqnarray}}
\newcommand\bea{\begin{eqnarray}}

\newcommand{\al}{$\alpha$}

\newcommand{\bsh}{\frac{\beta}{H_*}}
\newcommand{\bshd}{$\beta /H_*$}
\begin{document}
\draft
\input epsf
\twocolumn[\hsize\textwidth\columnwidth\hsize\csname
@twocolumnfalse\endcsname

\title{Supersymmetric Phase Transitions and Gravitational Waves at LISA}
\author{Riccardo Apreda${}^{(1)}$, 
Michele Maggiore${}^{(1)}$,
Alberto Nicolis${}^{(2)}$
and Antonio Riotto${}^{(2)}$}

\address{$^{(1)}${\it Dipartimento di Fisica, via Buonarroti 2, 
I-56100, Pisa and INFN, sezione di Pisa, Italy}}

\address{$^{(2)}${\it Scuola Normale Superiore, piazza dei 
Cavalieri, I-56125, Pisa and INFN, sezione di Pisa,  Italy}}

\date{February, 2001}
\maketitle
\begin{abstract}
Gravitational waves generated  during  a first-order  electroweak
  phase transition have a typical frequency which 
today falls just within  the band of the planned space interferometer
LISA. Contrary
to what happens in the Standard Model,  in its supersymmetric 
extensions 
the electroweak phase transition may be strongly first order, 
providing a mechanism for generating the observed baryon asymmetry
in the Universe. We show that during the same transition 
the
production of gravitational waves can be
rather  sizable. While the energy density in
gravitational waves  can reach  at most 
$h_0^2 \, \Omega_{\rm gw}\simeq 10^{-16}$ in the Minimal Supersymmetric
Standard Model,  in the Next-to-Minimal Supersymmetric
Model, in some parameter range,
$h_0^2 \; \Omega_{\rm gw}$ can  be as high as $4\times 10^{-11}$.
A  stochastic 
background of gravitational waves of this intensity
is within the reach of  the planned
sensitivity of LISA. Since in the Standard Model 
the  background of gravitational waves is totally neglegible, 
its detection  would also provide a
rather unexpected experimental signal of  supersymmetry and a tool
to descriminate among supersymmetric models with different Higgs content.

\end{abstract}
\pacs{PACS: 98.80.Cq;\ IFUP-TH/04-2001;\ 
SNS-PH-01-02,;\ hep-ph/yymmnn}
\vskip2pc]

\def\simlt{\stackrel{<}{{}_\sim}}
\def\simgt{\stackrel{>}{{}_\sim}}


During its evolution the Universe has probably undergone a series of 
 phase transitions and
some important remnants of these events may exist today, 
including the observed baryon asymmetry of the 
Universe and a stochastic background of gravitational waves (GWs).
During a first-order phase transition the Universe is ``trapped'' 
in a metastable state -- the false vacuum -- which is
separated from the true vacuum by a barrier in the potential
of the order parameter, usually a scalar field $\phi$. 
The transition      takes place through  the nucleation of
bubbles of the new phase and  most of the latent heat 
 released in the transition is converted
into kinetic energy that makes the bubbles expand. When  the bubble
 walls collide particle production
takes place \cite{coll} and part of the energy is radiated
into gravitational waves 
\cite{witten,hogan,thorne,TuWi,kosowsky,turner,envelope,kamionkowski}. 
The properties of 
this gravitational radiation -- such as the characteristic frequency and the 
intensity -- depend on the typical energy scales involved 
in the transition.
In particular, the electroweak phase transition 
is expected to produce a background of GWs
 with a peak frequency
around the milliHertz. This frequency  happens to be 
 the range most relevant for the space interferometer 
LISA~\cite{LISA}, which is planned to fly by 2010.

A strongly first order phase transition could  also
provide a mechanism for generating the observed baryon asymmetry
in the Universe~\cite{riotto}. 
For such a  reason, the strength of the phase transition
has been investigated in details 
in the  Standard Model (SM) and in extensions of it. 
Unfortunately, non-perturbative results have revealed that
there is no hot electroweak phase transition in the Standard Model
for Higgs masses  larger than $m_W$
\cite{pt}. Therefore, GWs are not produced at the SM electroweak
transition.

Among the possible extensions of the SM at the weak scale, its
supersymmetric extensions are the best motivated ones.
In the Minimal Supersymmetric Standard Model (MSSM),  a
 strong enough phase transition requires light Higgs and stop eigenstates. 
For a Higgs mass in the range $(110-115$) GeV, 
there is a window in the right-handed stop mass  $m_{\rm stop}$
in the range $(105-165)$ GeV \cite{reviewquiros}.
If the Higgs is heavier than about 115 GeV,
       stronger constraints are imposed on the space of 
supersymmetric parameters. However, the
strength of the transition can be further enhanced in extensions
of the MSSM, for instance with the addition of a gauge singlet in the 
Higgs sector \cite{pietroni}.

The goal  of this paper is to compute the
amount of gravitational waves  generated during the 
electroweak phase transition in supersymmetric extensions of the SM.
We will find that, depending on the model and on the region of
parameter space,
the stochastic background of GWs generated in the collision
of bubbles nucleated during the supersymmetric 
electroweak phase transition can be
within the sensitivity of the planned
space-interferometer LISA. This opens the exciting
possibility that the very same supersymmetric physics responsible for
the generation of the primordial baryon asymmetry is also able
to provide us with a detectable background of gravitational waves.

\begin{figure}
\centering\leavevmode\epsfxsize=2.8in\epsfbox{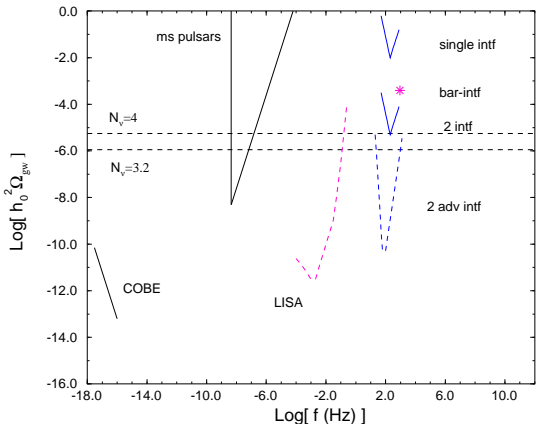}
\caption{The bounds from nucleosynthesis (horizontal dashed lines,
	for $N_{\nu}=4$ and for $N_{\nu}=3.2$), from COBE and from ms pulsars,
	together with the sensitivity of ground based detectors
	and of LISA. See ref.~[17]
        for details. \label{theo}}
\end{figure}

A stochastic background of
GWs~\cite{thorne,allen,maggiore} 
can be characterized by the dimensionless quantity
\be
\ogw (f)=\frac{1}{\rho_c}\,\frac{d\rho_{\rm gw}}{d\log f}\, ,
\ee
where $\rho_{\rm gw}$ is the energy density associated to GWs, 
$f$ is their frequency and $\rho_c$ is the present value of the
critical energy density, $\rho_c =3H_0^2/(8\pi G_N)$, with $H_0=100 h_0 
 $ Km/sec Mpc; $h_0$ parametrizes the uncertainty in $H_0$.
In fact, it is more convenient to characterize the stochastic  background
of GWs with the quantity $\hogw (f)$, which is independent of
$h_0$.
In Fig.~\ref{theo} we show the most relevant bounds on  
cosmological stochastic GW backgrounds, together with the experimental
sensitivities of the various detectors under construction, 
as discussed in Refs. \cite{maggiore,maggiorets}.
In particular, LISA is expected to reach a sensitivity of order
$\hogw \simeq  10^{-12}$ at $f=1$~mHz. At this frequency --
however -- a cosmological signal could be masked by  an astrophysical
background due to unresolved compact white dwarf binaries. Its
strength is uncertain, since it depends on the rate of white dwarf
mergers and it is estimated to be 
$\hogw \simeq  10^{-11}$~\cite{LISA}. At a frequency
$f\simeq 10$~mHz the LISA sensitivity is expected to be of order
$\hogw \simeq  10^{-11}$, and the astrophysical background is
below this value. 
(Correlating two detectors one usually gains
many orders of magnitudes in the sensitivity, but 
because of this extragalactic background,
even if one would fly two LISA detectors, the
sensitivity of the correlation would be limited to $\hogw \simeq 5\times
10^{-13}$~\cite{uv}).
 
Two are the basic quantities which play a role in the
determination of the GW background generated during a first-order
phase transition. The parameter  $\alpha$ 
gives a measure of the jump in the energy density 
experienced by the order parameter $\phi$
during the   
transition from the false to the true vacuum and  it is
the ratio between the false vacuum energy density 
and the energy density of the radiation at the transition temperature $T_*$.
 The parameter 
$\beta$ characterizes the bubble nucleation rate per unit volume, which 
can be expressed as $\Gamma=\Gamma_0 \exp(\beta t)$ \cite{turner}. 
Thus $\beta ^{-1}=\Gamma/\dot{\Gamma}$ represents roughly 
the duration of the phase transition and   the characteristic frequency of 
the 
gravitational radiation is expected to be $2\pi f \simeq \beta$. 
The stronger is the transition, the larger is $\alpha$ and the smaller
is $\beta$ because of the
larger amount of supercooling experienced by the system before the
transition. 
The present ({\it i.e.}  properly red-shifted) 
peak frequency of the GW background is determined by $\beta$ and by
$H_*$, $T_*$, $g_*$, which are respectively the Hubble parameter, 
the temperature and the number of relativistic degrees of freedom 
at the moment of the transition
\cite{kamionkowski},
\be
\fp \simeq 5.2 \times 10^{-8} \:  \(\frac{\beta}{H_*} \) \(\frac{T_*}{1\: {\rm GeV}} \) \( \frac{g_*}{100}\)^{1/6}  \: {\rm Hz}\:. \label{fpicco}
\ee 
Typical values for $\beta / H_*$ for the electroweak phase transition 
are  between $10^2$ and $10^3$, with $T_*={\cal O}(100)$ GeV. This gives
a  frequency $\fp$ in the range $(10^{-4}- 5 \times 10^{-3})$ Hz. 
From Fig. ~\ref{theo} we see that this is precisely the range 
in which LISA achieves its maximum sensitivity.

The intensity of the radiation produced is given by \cite{kamionkowski} 
\begin{eqnarray}
\hogw & \simeq &  10^{-6} \(\frac{0.7 \: \alpha + 
0.2 \sqrt{\alpha}}{1+0.7 \: \alpha} 
\times \frac{\alpha}{1+\alpha}\)^2  \times       \nonumber \\
 & & \times \(\frac{H_*}{\beta}\)^2
\(\frac{v^3}{0.25+v^3}\)\(\frac{100}{g_*}\)^{1/3}\, ,		\label{omega}
\end{eqnarray}
where $v$ is the velocity at which the bubble 
expands. In the following we will use a value of the velocity $v=v(\alpha)$
as given in Ref. \cite{kamionkowski} for bubble detonation.
Eq.~(\ref{omega}) makes it clear  that -- in order to produce a 
relevant signal -- one needs large $\alpha$ and 
small $\beta$, {\it i.e.} a  strongly first-order transition.

Before launching ourselves into details, let us   briefly describe 
how we have computed the 
relevant quantities appearing in Eqs.~(\ref{fpicco}) and (\ref{omega}), 
namely $T_*$, \al\ and $\beta / H_*$, once the thermal effective potential 
$V(\phi,T)$
for the Higgs scalar field(s) is given. 
The rate of tunneling per unit volume from the metastable minimum to 
the stable one is suppressed by the exponential of an effective 
action $
\Gamma = \Gamma_0 \; e^{- \frac{S_3(T)}{T}}\;  \label{gamma}$, 
where $\Gamma_0$ is of the order of $T^4$, and $S_3$ is the 
extremum of  the spatial Euclidean action $
S_3(T)=\int d^3 x \[ \half  (\vec{\nabla} \phi_b)^2 + V(\phi_b,T)  \] $
computed for the configuration of the scalar field(s) describing the
bubble wall which interpolates between the false and the true vacuum.
The transition takes place when the probability of nucleating a single 
bubble within one horizon volume becomes ${\cal O}(1)$.  This 
condition fixes $T_*$ to satisfy $S_3(T_*)/T_* 
\simeq \ln\; (\mpl/100\; {\rm GeV})^4\simeq 140$ 
for the electroweak 
transition. The parameter  \al \  
is readily computed from the 
definition given above, while \bshd \  
is given by
\be
\label{def}
\bsh = \left. T_*\;\frac{d \( S_3/T \)}{d T} \right|_{T_*}.
\ee

As we have already noticed, extensions of the SM are required to 
obtain a first-order phase transition at the electroweak scale.

In the MSSM two complex Higgs doublets $H_1$ and $H_2$ 
are present in the Higgs sector and new CP-violating phases appear which can 
drive enough amount of baryon asymmetry.  If the mass $m_A$ of the CP-odd
field in the Higgs sector  is much  larger than $m_W$, 
only one light Higgs scalar $\phi$ is left
 and its potential is similar to the one in the SM. When $m_A\sim m_W$
the full two-Higgs potential should be considered, but the strength of
the phase transition is weakened \cite{reviewquiros}.
The 
one-loop thermal 
corrections to the effective potential make the quadratic term
 positive at high temperature and create a negative
cubic term due to loops of the massive bosons in the theory. Due 
to the presence of this cubic term and the positivity of the
quadratic one, there exists a range of temperature in which the 
point $\phi = 0$ is a local minimum
separated from the true simmetry-breaking one by a small potential barrier,
 that is precisely 
the set-up for a first-order phase transition.  The 
strength of the  phase transition is enhanced
by the presence of  new bosons coupled to the Higgs, a significant
role being played by the right-handed  stop, which is 
-- apart from the Higgs itself --   the
lightest scalar  in the theory and has 
the largest Yukawa coupling to the Higgs $\phi$.

To study the amount of gravitational waves generated during the electroweak
phase transition within the MSSM  we have made use
of the thermal potential corrected up to two-loop level.
Indeed, two-loop corrections have been shown to render the phase transition
significantly stronger in the MSSM \cite{reviewquiros}. 
The most relevant parameters in the game  are the Higgs mass 
$m_{\rm higgs}$,
 the right-handed stop mass $m_{\rm stop}$
 and the zero temperature ratio between the vacuum expectation values
of the two neutral 
Higgses $\tan\beta_{{\rm MSSM}}=\langle H_2\rangle/\langle H_1
 \rangle$. 

\begin{figure}
\centering\leavevmode\epsfxsize=2.8in\epsfbox{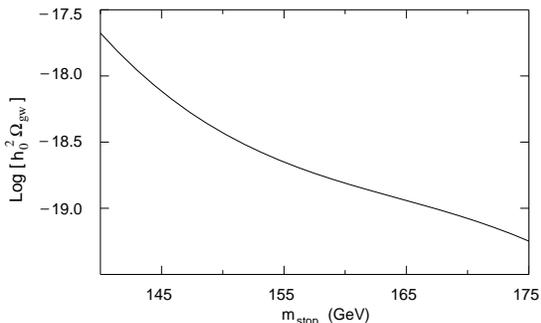}
\caption{ $\hogw$ as a function of the stop mass for a 110 GeV Higgs mass
. \label{MSSMmh}}
\end{figure}

\begin{figure}
\centering\leavevmode\epsfxsize=2.8in\epsfbox{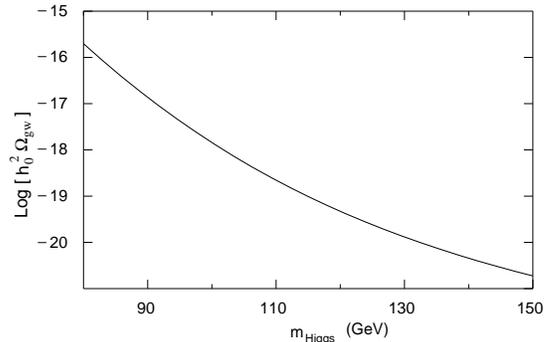}
\caption{$\hogw$ as a function of the Higgs mass for a 155 GeV stop
 mass
. \label{MSSMms}}
\end{figure}

Our strategy has been the following. 
For any given choice of the parameters of the model, we have 
first numerically computed the nucleation temperature $T_*$ by imposing that, for the
Higgs field configuration describing the nucleated bubble, 
the condition $S_3(T_*)/ T_* \simeq 140$ is satisfied. Then we have computed the parameters $\alpha$ and  $\beta$ through  Eq.~(\ref{def}). Our results are summarized in Figs.~\ref{MSSMmh} and \ref{MSSMms}.

The general prediction is that the intensity of the
gravitational waves produced during the MSSM phase transition
is too small for LISA. For instance, 
taking a Higgs mass of  110 GeV, the right-handed 
stop mass of 140 GeV and $\sin^2 \beta_{{\rm MSSM}} = 0.8$, we find 
$\alpha \simeq 3\times 10^{-2}$ and \bshd $\simeq 4\times 10^3$, leading
 to $\hogw \simeq 2 \times
10^{-18}$. 
Notice that one is not allowed to lower too much the
right-handed  stop mass   because the thermal squared mass 
for the stop itself would become negative at small temperature, 
thus leading to a physically unacceptable stable 
color breaking vacuum state at zero 
temperature.

We have also estimated  what happens if we lower
the Higgs mass down to (the already excluded value of) 80 GeV, 
setting  the right-handed stop mass at 155 GeV -- which is 
the lower value compatible with the 
absence of color breaking minima -- and $\sin^2 \beta = 0.8$. We obtain  
$\alpha \simeq 0.1$ and \bshd $\simeq 2\times 10^3$, giving 
$\hogw \simeq 2 \times
10^{-16}$, a signal still not relevant. The situation does
not  improve  when both Higgses are involved in the transition because
the strength of the phase transition is weaker.
A complete analysis of the results
within the MSSM will be presented elsewhere \cite{inprep}.
An uncertainty in this estimate is due to the determination of $v$. If
the phase transition is not strong enough, then
$v$ is subsonic, so that the value of $\hogw$
that we have found is really an upper bound.
 
The situation improves considerably if we enlarge the MSSM sector
adding a gauge singlet $N$ \cite{ellis}. This is the so-called Next-to-Minimal
Supersymmetric Standard Model (NMSSM) and 
is a 
particularly 
attractive model  to explain the observed baryon asymmetry
at the electroweak phase transition. 
The relevant part of the superpotential
is given by $
W=\lambda H_1 H_2 N-\frac{k}{3} N^3$,
where now the supersymmetric $\mu$-parameter of the MSSM is 
substituted by the the combination
$\lambda \langle N\rangle$, and $k$ is a free parameter. 
The corresponding Higgs potential reads
$V=V_{F}+V_{D}+V_{{\rm soft}}$, where 
\bea \label{pot}
V_{F} & = & |\lambda|^{2}\left[|N|^{2}(|H_{1}|^{2}+|H_{2}|^{2})
+|H_{1}H_{2}|^{2}\right]
       \nonumber \\ 
                &+& k^{2} |N|^{4} -(\lambda k^{*} H_{1}H_{2}{N^{2}}^{*} + 
{\rm h.c.})\, ,
\nonumber \\ \vspace{0.5 truecm}
V_{D} & = & \frac{g^{2}+g'^{2}}{8} (|H_{2}|^{2}-|H_{1}|^{2})^{2} +
\frac{g^{2}}{2}|H_{1}^{\dag}H_{2}|^{2}\, ,\nonumber \\ \vspace{0.5 truecm}
V_{{\rm soft}} & = & m_{H_{1}}^{2}|H_{1}|^{2} + m_{H_{2}}^{2}|H_{2}|^{2} + m_{N}^{2}
|N|^{2} \nonumber \\ & - & \left(\lambda A_{\lambda}H_{1}H_{2}N
-\frac {1}{3} k A_{k} N^{3} + {\rm h.c.}\right). 
\eea
The presence of the cubic supersymmetry breaking soft terms 
proportional  to the parameters  $A_\lambda$ and $A_k$
 already at zero temperature makes it clear that
within the NMSSM it is quite easy to get a very strong first-order
phase transition at the electroweak scale \cite{pietroni}.
The order of the transition is determined by these trilinear soft 
       terms rather than by the cubic term appearing in the
finite temperature one-loop corrections and the
 preservation of baryon
       asymmetry after the phase transition is
possible  for masses of the lightest scalar up to about 170 GeV. 
Barring the possibility that the transition occurs along CP-violating 
directions, the potential becomes a function of three real scalar fields
$({\rm Re } N, {\rm Re} H_1, {\rm Re} H_2)$. 
In our numerical analysis we have made use of 
 the tree-level potential (\ref{pot}) plus
the one-loop corrections appearing at finite temperature.
Overall, we have   six free parameters: the coupling parameters $\lambda$
and $k$; the soft-breaking mass terms  $A_{\lambda}$ and  $A_k$;
 the zero-temperature vacuum expectation value  of  the singlet $x$
 and $\tan\beta_{{\rm MSSM}}$.

Given a set of parameters, the strategy has been again to determine
the nucleation temperature $T_*$ and the parameters
$\alpha$ and $\beta$ which in turn determine the intensity and the
frequency of the stochastic gravitational background.
In our analysis we have focussed on those regions of the
parameter space which previous studies have shown to 
give rise to a large enough baryon asymmetry.

\begin{figure}
\centering\leavevmode\epsfxsize=2.8in\epsfbox{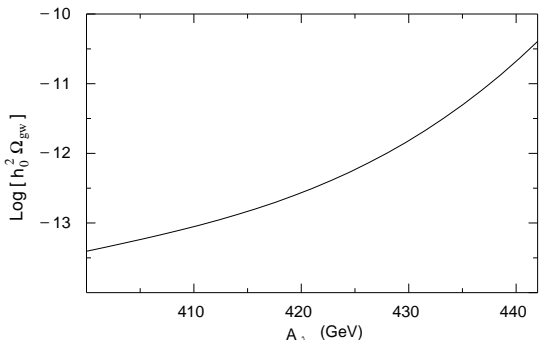}
\caption{$\hogw$ as a function of $A_\lambda$ for $A_k=480$ GeV, $x=350$ GeV, 
$\lambda=0.83$, $k=0.67$ and $\tan \beta_{{\rm MSSM}}=2$
 \label{NMSSMak}}
\end{figure}

\begin{figure}
\centering\leavevmode\epsfxsize=2.8in\epsfbox{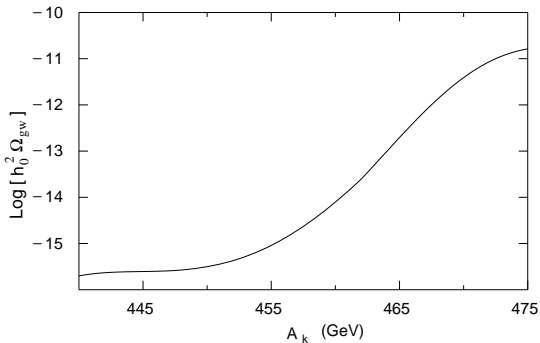}
\caption{$\hogw$ as a function of $A_k$ for $A_\lambda=450$ GeV, $x=350$ GeV, 
$\lambda=0.83$, $k=0.67$ and $\tan \beta_{{\rm MSSM}} =2$
 \label{NMSSMal}}
\end{figure}

Typical  results are summarized in Figs.~\ref{NMSSMak} and \ref{NMSSMal}, 
where we plot $\hogw$ as a function of $A_\lambda$ and $A_k$, respectively,
with $A_k=480$ GeV in Fig.~\ref{NMSSMak} and
$A_\lambda=450$ GeV in Fig.~\ref{NMSSMal}. Note also that, since the
transition in the NMSSM is strongly first order,  it is correct to
use the expression for $v(\alpha )$ computed for detonation 
bubbles~\cite{kamionkowski}.

The intensity of the produced gravity waves rapidly grows as a function
of the soft breaking mass terms. This is explained by the fact that
$\hogw$ scales roughly as  $\alpha^3/\beta^2$ and a mild increase of 
$\alpha$ and decrease of $\beta$ leads to a rapid growth of the
intensity. However, $A_\lambda,A_k$ cannot be increased beyond the
values shown in the figure because for higher values 
the condition $S_3(T_*)/ T_* \simeq 140$ is never satisfied and the
transition does not take place.

The conclusion is that values
 $\hogw\simeq  4\times 10^{-11}$ are
reachable in the NMSSM in some regions of the parameter space
(we have found similar values of $\hogw$
also in other regions of the parameter space~\cite{inprep}).
A background of this intensity would be within the
reach of LISA. 

We find interesting that the very same supersymmetric 
phase transition which
provides us with  a mechanism to generate the observed baryon asymmetry
might also be the source of  a sizeable background
of gravitational waves. If such background is detected, it will be not only 
an indication that supersymmetry might play a role at the electroweak phase
transition, but even a way to discriminate among supersymmetric
models with different Higgs sectors.
%
%

\end{document}